\documentclass{ws-jbcb}
\usepackage{
   amsfonts,
   xspace,
   colortbl,
   graphicx}

\newcommand{\ds}         
  {\displaystyle}

\newcommand{\fns}        
  {\footnotesize}

\newcommand{\llv}        
  {\left| \left|}

\newcommand{\rrv}        
  {\right| \right|}

\newcommand{\bdb }       
  {\mbox{\boldmath $\beta$}}

\newcommand{\bdw }       
  {\mbox{\boldmath $w$}}

\newcommand{\bdx }       
  {\mbox{\boldmath $x$}}

\newcommand{\bdy }       
  {\mbox{\boldmath $y$}}

\newcommand{\bd }[1]     
  {\mbox{\boldmath $#1$}}

\newcommand{\hbd }[1]    
  {\hat{\mbox{\boldmath $#1$}}}

\newcommand{\spc}        
  {\,\,\,\, }

\newcommand{\R}          
  {{\mathbb R} }

\newcommand{\ones}[1]    
  {\mbox{\boldmath $1$}_{#1}}

\newcommand{\zeros}[1]   
  {\mbox{\boldmath $0$}_{#1}}

\newcommand{\txtsub}[2]
  { \mbox{$#1_{\mbox{\tiny #2}}$} }

\newcommand{\iie}
  {\emph{i.e.$\,$}}

\newcommand{\etal}
  {\emph{et al.}}

\newcommand{\ieg}
  {\emph{e.g.$\,$}}

\begin{document}
%
\markboth{Andries E, Hagstrom TM, Atlas SR, Willman C}
         {REGULARIZATION STRATEGIES FOR HYPERPLANE CLASSIFIERS}
\catchline{}{}{}{}{}

\title{REGULARIZATION STRATEGIES FOR HYPERPLANE CLASSIFIERS:
APPLICATION TO CANCER CLASSIFICATION WITH GENE EXPRESSION DATA}

\author{ERIK ANDRIES}
\address{Departments of Mathematics \& Statistics and Pathology\\
        University of New Mexico, Albuquerque, NM 87131, USA\\
        andriese@math.unm.edu}

\author{THOMAS HAGSTROM}
\address{Department of Mathematics \& Statistics\\
         University of New Mexico, Albuquerque, NM 87131, USA\\
         hagstrom@math.unm.edu}

\author{SUSAN R. ATLAS}
\address{Center for Advanced Studies and Department of Physics \& Astronomy\\
         University of New Mexico, Albuquerque, NM 87131, USA\\
         susie@sapphire.phys.unm.edu}

\author{CHERYL WILLMAN}
\address{Cancer Research and Treatment Center and Department of Pathology\\
         University of New Mexico, Albuquerque, NM 87131, USA\\
         cwillman@salud.unm.edu}
\maketitle
\begin{history}
\received{(Day Month Year)}
\revised{(Day Month Year)}  
\accepted{(Day Month Year)}
\end{history}  
%
\begin{abstract}
Linear discrimination, from the point of view of numerical linear algebra, 
can be treated as solving an ill-posed system of linear 
equations.  In order to generate a solution that is 
robust 
in the presence of noise, these problems require regularization. 
Here, we examine the ill-posedness involved in the
linear discrimination of cancer gene expression data with respect to  
outcome and tumor subclasses.  
We show that a filter factor representation, 
based upon Singular Value 
Decomposition, yields 
insight into the numerical ill-posedness 
of the hyperplane-based separation when applied  
to gene expression data.  We also show that this
representation yields useful diagnostic
tools for guiding the selection of classifier parameters, 
thus leading to improved performance. 
\end{abstract}
\newpage
%
%
\section{Introduction}
%

A current challenge in cancer treatment is to target specific 
therapies to distinct tumor types and to tailor the intensity of 
the treatment to the risk of relapse for each 
patient\cite{carroll,veer,willman}. 
Crucial to this effort is to effectively classify patients into 
specific risk groups.  Traditionally, many risk stratification schemes
use tumor morphology, molecular genetics and cytogenetics in addition to 
utilizing information such as race, age, etc.\cite{who}
Clinically, 
however, this approach has limitations. Tumors with similar 
appearances may have different clinical courses and display 
different responses to therapy.  In addition, conventional 
laboratory diagnostic procedures do not reveal the full underlying 
molecular heterogeneity of these tumors. For this reason, there has 
been great interest in using gene arrays to identify more precisely 
known tumor subclasses, to discover new tumor subclasses, and to 
predict {\em outcome}, \iie whether or not a 
patient's cancer will go into remission.

Our goal in this work is to investigate the classification of
cancer patient samples according to tumor lineage and outcome
via hyperplane classifiers, \iie we attempt to linearly separate
with a hyperplane
patient samples via their expression profiles.
This is numerically challenging since gene expression 
data is characterized by a
noisy, high-dimensional, low-sample-size setting.
Hyperplane classifiers cope with this challenging setting
by incorporating prior knowledge assumptions that constrain
the solution (the hyperplane parameters)
in some way.
The term {\em regularization} refers
to this incorporation of prior
information in order to stabilize the problem and to sift
out a desired solution.
With respect to hyperplane classifiers, there are many choices
but each one encodes a different
strategy for how to stabilize the solution, and,
in this paper, we focus on the
{\em filter factor representation} using singular value
decomposition (SVD) as the
particular regularization strategy.
While some researchers have used SVD-based
approaches\cite{ghosh_svd,nguyen} for cancer classification,
to the best of our knowledge, there has been no work on 
differentiating these methods on the basis of a 
filter factor representation.
In addition, in seeking to evaluate
why a certain regularization strategy should be preferred over
another in the context of gene expression data,
the type of numerical ill-posedness (\emph{rank-deficiency}
or \emph{discrete ill-posedness})
exhibited by expression data ought to be taken into account.
To date, this has not been done.  Moreover, the filter 
factor representation involves the use of 
\emph{spectral coefficients} that can be used to estimate how
much regularization should be applied. 

In Section \ref{sec:notation}, the mathematical
notation used in subsequent sections will be detailed.
In Section \ref{sec:filfacrep}, we discuss kernel versions
of regression-based hyperplane classifiers and the
regularization
approaches that these hyperplane classifiers can adopt when
using a SVD-based filter factor representation.
In Section \ref{sec:data}, we review the cancer
expression data sets used in our analysis
and the methods used for data
preprocessing and parameter estimation.
Section \ref{sec:res} presents the classification results
for hyperplane classifiers across a variety of regularization
schemes and describes how spectral coefficients such as the singular values
and Fourier coefficients can be used to tune classifier parameters.  Finally, 
we present the conclusion in \ref{sec:concl}.

\section{Notation} 
\label{sec:notation} 

The matrix $\bd{X} = [\bdx_1,\ldots,\bdx_n] \in \R^{d \times n}$ denotes 
the gene expression data derived from $n$ patient tissue samples with 
each patient expression vector $\bd{x}_i \in \R^d$ consisting of $d$ 
gene attributes. The vector $\bd{y} \in \R^n$ consists of $n$ 
binary-valued class labels where $y_i =\{-1,+1\}$ indicates membership 
in either the negative or positive class.  With respect to acute 
leukemias, for example, one might assign the class label $y_i=-1$ or 
$y_i=+1$ to indicate membership in either ALL (acute lymphoblastic 
leukemia) or AML (acute myeloid leukemia), respectively.  The matrix 
$\bd{Y}$ is an $n \times n$ diagonal matrix such that $\bd{Y} = 
\mbox{diag}(\bdy) = \mbox{diag}(y_1,\ldots,y_n)$ and $\bd{Y}^2 = 
\bd{I}_n$.  The vectors $\ones{n}$ and $\zeros{n}$ represent column 
vectors of $n$ ones and zeros, respectively.  The transpose of a matrix 
or vector is denoted by the superscript $T$, while $\|\cdot\|_2$ 
denotes the two-norm of a vector. The superscript $p$ will refer to a 
specific partition of patient samples into a training and test set (this 
will be explained further in \S \ref{sec:res}).

Given $\bd{X}$ and $\bd{y}$, one attempts to learn the coefficients, 
$\bd{w} \in \R^d$ and $w_0$, that determine a separating hyperplane 
given by the following mathematical expression:  
$g(\bdx) = \bdx^T \bdw + w_0 \,=\, 0$
\cite{cristianini}.
Geometrically, different choices of $\bdw$ and $w_0$ typically yield
different hyperplanes since $\bd{w}$ determines the tilt of the hyperplane,  
while $w_0$ is the bias, or offset from the origin. To simplify notation, 
the data is often augmented to incorporate $w_0$:
$\hbd{w}=[\bd{w}^T,\, w_0]^T \in \R^{d+1}$
and $\hbd{x} = [\bd{x}^T, \, 1]^T \in \R^{d+1}$ denote augmented
weight and gene expression vectors, respectively.  
Hence, $g(\bdx) = \bdx^T \bdw + w_0$ can be simply expressed as 
$g(\bdx) = \hbd{x}^T \hbd{w}$.
Similarly,
$\hbd{X}=[\bd{X}^T, \ones{n}]^T \in \R^{(d+1) \times n}$ is the 
augmented data for the entire gene expression matrix.
This augmentation adds a constant feature to the training
data such that the separating hyperplane passes through the origin in
$\R^{d+1}$. The algorithm that determines the optimal hyperplane parameters
$\hbd{w}$ will be referred to as a {\em hyperplane classifier}, and the 
data
$\bd{X}$ and $\bd{y}$ used to determine $\hbd{w}$ is referred to as
the {\em training data}.
Once $\hbd{w}$ is obtained,
the formal classification task is to assign class membership to a new input 
$\bd{z} \in \R^d$: if
$g(\bd{z}) < 0$ ($g(\bd{z}) > 0$), then $\bd{z}$ is assigned to
the negative (positive) class.  If $g(\bd{z}) = 0$, then the class
membership of $\bd{z}$ is indeterminate.
If
$\hbd{Z} = [\hbd{z}_1,\ldots,\hbd{z}_N] \in \R^{(d+1) \times N}$
consists of $N$ augmented test set
expression profiles distinct from the training set and
$\bd{t} \in \R^N$ is its corresponding
set of class labels such that $t_i = \{-1,+1\}$, then
the number of misclassifications $M$ can be
computed as follows:
\begin{equation}
  M = \sum_{i=1}^N I[t_i g(\hbd{z}_i)]
           = \sum_{i=1}^N I[t_i (\hbd{z}_i^T \hbd{w})]
  \label{eq:misclass}
\end{equation}
where $I[\cdot]$ is the indicator function such that 
$I[\theta]= 1$ if $\theta < 0$ and $I[\theta] = 0$
if $\theta \geq 0$.

%
\section{Filter Factor Representations of Hyperplane Classifiers}
\label{sec:filfacrep}
%

A hyperplane classifier, in its simplest form, specifies
a linear relationship between a response variable, $\bdy$, and a
set of predictor variables, $\bd{X}$. For many problems, estimates of the
linear relationships between variables are adequate to describe the
observed data and to make reasonable predictions for new observations.
This has been well-established in the case of
cancer classification using gene expression
data.\cite{ghosh_svd,kang,khan,nguyen} 
However, since the number of samples ($n$) is much less than the 
number of genes ($d$), the samples sparsely populate a very high
dimensional gene space and, as a result, there is a strong likelihood of 
finding many perfectly separating hyperplanes for the training data.
Another complicating factor is the presence of significant biological 
and experimental variability in the expression data. 
Assuming that a linear discrimination approach to the classification 
of gene expression data will generally suffice, we therefore 
focus on deciding which regularization strategy is 
appropriate for extracting a \emph{stable} set of
hyperplane parameters in the presence of 
noisy and high-dimensional data.

\subsection{Kernel-based regression}
Suppose that the hyperplane classifier attempts to find 
$\hbd{w}$ by solving the regression
problem $\hbd{X}^T \hbd{w} = \bdy$ using
least squares (LS) minimization:
\begin{equation}
  {\ds
  \min_{\hat{w}} \,\, \llv \hbd{X}^T \hbd{w} - \bdy \rrv_2^2
  \, = \,
  \min_{\hat{w}} \, \sum_{i=1}^n (\hbd{x}_i^T\hbd{w}-y_i)^2
  \, = \,
  \min_{\hat{w}} \, \sum_{i=1}^n (g(\hbd{x}_i)-y_i)^2
  }
  \label{eq:regression}
\end{equation}
LS methods for solving for the weight vector $\hbd{w}$ 
(hereafter referred
to as the {\em primal variable}), 
such as QR factorization, involve $O(dn^2)$ floating point 
operations ({\em flops})\cite{demmel}.
When $d \gg n$, one can reduce the computational complexity
by working with the 
kernel version of Eq.(\ref{eq:regression}) instead.  
Assuming that $\hbd{w}$ can be rewritten as a linear
combination of the training data, \iie,
$\hbd{w} = \hbd{X} \bdb$, $\bdb \in \R^n$,
$\bdb$ (hereafter referred to as the {\em dual variable}) 
can be determined instead:
\begin{equation}
  {\ds
  \min_{\hat{w}} \, \llv \hbd{X}^T \hbd{w} - \bdy \rrv_2^2
  \, = \,
  \min_{\beta}   \, \llv \hbd{X}^T (\hbd{X}\bdb) - \bdy \rrv_2^2
  \, = \,
  \min_{\beta}   \, \| \bd{K} \bdb - \bdy \|_2^2,
  }
  \label{eq:kernel_regression}
\end{equation}
where $\bd{K}=\hbd{X}^T \hbd{X} \in \R^{n \times n}$ will be
referred to as the linear {\em kernel} matrix.  Solving
Eq.(\ref{eq:kernel_regression})
now scales cubically with
the number of patient samples $n$.

\subsection{Need for regularization}

If $\hbd{X}$ is ill-conditioned, then the computed solution $\bdb$ will 
not be stable.  Ill-conditioning is always the case with coefficient 
matrices derived from classification-based regression problems. For 
example, if $\hbd{X}=[\hbd{X}_{+}, \, \hbd{X}_{-}]$ is a data partitioning
with respect to the positive and negative classes, then the 
feature or attribute profiles of 
data points 
within $\hbd{X}_{+}$ and $\hbd{X}_{-}$ will be highly correlated
because it is this commonality which defines the class.  As a 
result, the effective numerical rank $\txtsub{r}{eff}$ (the number of 
columns of $\hbd{X}$ that are linearly independent with respect to some 
error level) will be small due to the multicollinearity of the columns 
within $\hbd{X}_{+}$ and $\hbd{X}_{-}$. In the case of gene expression data, 
$\txtsub{r}{obs}$ can vary significantly due to the large amount of 
within-class and between-class biological variability.  In addition, the 
large amount of noise associated with gene arrays can artificially 
inflate $\txtsub{r}{eff}$ such that the observed rank $\txtsub{r}{obs}$ 
will be greater than $\txtsub{r}{eff}$. In general, $\txtsub{r}{obs}=n$ 
(full rank) but $\txtsub{r}{eff} \ll \txtsub{r}{obs}$.

The phenomenon of solution instability can be illustrated by
expressing the ordinary least squares (OLS) solution to
Eq.(\ref{eq:kernel_regression}) in terms
of the singular value decomposition (SVD) of
$\hbd{X}$\cite{demmel,trefethen}.  
If $\hbd{X} = \bd{U \Sigma V}^T$
is the SVD of $\hbd{X}$, where
\[ \begin{array}{c}
  \bd{U}=[\bd{u}_1,\ldots,\bd{u}_n] \in \R^{(d+1) \times n}, \spc
  \bd{V}=[\bd{v}_1,\ldots,\bd{v}_n] \in \R^{n \times n}, \\
  \bd{U}^T \bd{U} = \bd{V}^T \bd{V} = \bd{I}_{n}, \\
  \bd{\Sigma}=\mbox{diag}(\sigma_1,\ldots,\sigma_n), \,\,
  \sigma_1 \geq \cdots \geq \sigma_n \geq 0, 
\end{array} \]
then the solution $\bdb$ can be given in terms of
$\bd{\Sigma}$ and $\bd{V}$:
\begin{equation}
  \bdb \,=\, \bd{K}^{-1}\bdy
       \,=\, \bd{V} \bd{\Sigma}^{-2} \bd{V}^T \bdy
       \,=\, \sum_{i=1}^n \left(
             \frac{ \bd{v}_i^T \bdy }{ \sigma_i^2 }
             \right) \bd{v}_i.
  \label{eq:ols_soln}
\end{equation}
Since we assumed $\hbd{X}$ to have full rank
(even though it is ill-conditioned), we use
$\bd{K}^{-1}$ instead of replacing it with the 
pseudoinverse $\bd{K}^{\dagger}$.
In Eq.(\ref{eq:ols_soln}), the terms 
associated with small singular values (\iie, terms associated with large 
$i$) correspond to spurious noise inherent in the data.  Division by 
small singular values unduly amplifies the effect of these noise terms 
and has a direct analogue with overfitting: if we include terms 
associated with small singular values, then the solution will have low 
{\em bias} and high {\em variance}\cite{hastie}.  
Numerically, this overfitting will 
manifest itself as a large solution norm for $\bdb$ since $\| \bdb 
\|_2^2 = \sum_{i=1}^n [(\bd{v}_i^T \bdy)/\sigma_i^2]^2$.  
Regularization in the context of the SVD expansion of 
Eq.(\ref{eq:ols_soln}) amounts to eliminating or damping noise terms 
associated with small singular values.  This, in effect, imposes a prior 
knowledge assumption of ``smoothness'' on the solution since large 
values of $\| \bdb \|_2^2$ are penalized.

\subsection{Filter factor representations}
There are a variety of regularization techniques 
that can be used to impose
solution smoothness.  We are particularly interested in techniques
that admit {\em filter factor solutions}\cite{hansen}.
A filter factor representation is a reweighting
of Eq.(\ref{eq:ols_soln}) and has the form
\begin{equation}
    \bdb
    \,=\, \bd{VF} \bd{\Sigma}^{-2} \bd{V}^T \bdy
    \,=\, \sum_{i=1}^n \left(
          \frac{ \bd{v}_i^T \bdy }{ \sigma_i^2 }
          \right) f_i \bd{v}_i,
    \label{eq:ff_soln}
\end{equation}
where $\bd{F}=\mbox{diag}(f_1,\ldots,f_n)$ is a diagonal matrix 
consisting of the {\em filter factors}\cite{hansen} or {\em shrinkage 
coefficients}\cite{hastie}. The filter factors typically decay to zero 
as $i$ increases such that contributions from terms with small singular 
values are filtered out.  For the OLS solution, no damping occurs since 
$\bd{F}=\bd{I}_n$ (or $f_1=\cdots=f_n=1)$.  However, least squares (LS) 
techniques such as truncated singular value decomposition (TSVD), kernel 
ridge regression (KRR) and partial least squares (PLS) have nontrivial 
filter factor representations as a result of solving a modified LS 
problem involving the dual variable $\bdb$.  
Each of these LS methods, through its respective filter factor
representation,
encodes a different regularization strategy in order 
to suppress the noisy 
terms in the SVD expansion of Eq.(\ref{eq:ols_soln}).
We briefly describe
each of these methods in turn. 

\subsubsection{TSVD}
\label{sec:tsvd_ff}
For TSVD, the approach to damping noise terms in
Eq.(\ref{eq:ols_soln}) is to simply eliminate
them entirely.  Keeping the first $k$ terms amounts
to having binary-valued filter factors such that
$f_1=\cdots=f_k=1$ and $f_{k+1}=\cdots=f_n=0$.
Alternatively, one can arrive at the same solution
by solving the modified LS problem, $\bd{K}_k \bdb = \bdy$,
in which
$\bd{K}_k=\sum_{i=1}^k \sigma_i \bd{v}_i \bd{v}_i^T$ is
the rank-$k$ approximation of $\bd{K}$ obtained by
replacing the smallest $n-k$ singular values with zeros.

\subsubsection{KRR}
\label{sec:krr_ff}
For KRR, the modified LS problem involving $\bdb$
is derived by minimizing the following loss function,
$L(\hbd{w}) = \| \hbd{X}^T \hbd{w}-\bdy \|_2^2
 + \lambda^2 \| \hbd{w} \|_2^2$,
by setting $\nabla_{\hat{w}} L(\hbd{w})=\zeros{d+1}$ and
replacing $\hbd{w}$ with $\hbd{X} \bd{\beta}$:
\begin{equation}
  \nabla_{\hat{w}} L(\hbd{w}) =
  2(\hbd{X} \hbd{X}^T + 
  \lambda^2 \bd{I}_{d+1})\hbd{w} - 2\hbd{X} \bdy =
  \zeros{d+1}
  \,\, \Rightarrow \,\,
  (\bd{K} + \lambda^2 \bd{I}_{n})\bdb = \bdy.
  \label{eq:krr}
\end{equation}
Note, however, that the linear system in
Eq.(\ref{eq:krr}) involving $\bdb$
is not of the form typically associated with classical
ridge regression or Tikhonov regularization,
\begin{equation}
  (\bd{A}^T \bd{A} + \gamma^2 \bd{B}^T \bd{B})\bdx = \bd{A}^T \bd{b},
  \label{eq:tikh_reg_system}
\end{equation}
in which the matrices $\bd{A} \in \R^{m \times n}$ and
$\bd{B} \in \R^{n \times p}$ $(m \geq n \geq p)$ are not assumed to be
symmetric positive definite (SPD).
Instead, $(\bd{K} + \lambda^2 \bd{I}_{n})\bdb = \bdy$
corresponds to a related regularization scheme first proposed by
Franklin\cite{franklin} (hereafter referred to as the {\em Franklin}
regularization scheme) in which he suggested
replacing Eq.(\ref{eq:tikh_reg_system})
with
\begin{equation}
  (\bd{A}+\gamma \bd{B})\bdx=\bd{b},\, \gamma > 0,
  \label{eq:franklin_reg_scheme}
\end{equation}
when $\bd{A}\in \R^{n \times n}$ and $\bd{B} \in \R^{n \times n}$
are SPD.
If $\bd{B}=\bd{I}_n$ and
$\{\psi_1,\ldots,\psi_n\}$ are the singular values of $\bd{A}$, then
the filter factors are the following\cite{hansen} :
\[
  f_i = 
  \left\{ \begin{array}{ll}
  \psi_i^2/(\psi_i^2+\gamma^2), & \mbox{Tikhonov regularization}, \\ 
  \psi_i/(\psi_i+\gamma),       & \mbox{Franklin regularization}.
  \end{array} \right. 
\]
Hence, KRR,
in the context of Eq.(\ref{eq:tikh_reg_system}),
is nothing more than the Franklin regularization scheme with
$\bd{A} \equiv \bd{K}$ (where $\psi_i=\sigma_i^2$),
$\bd{B}\equiv\bd{I}_n$, $\bdx\equiv\bdb$, $\bd{b}\equiv \bdy$ and
$\gamma \equiv \lambda^2$.

In the context of support vector machine (SVM) 
classification\cite{cristianini},
KRR also has connections with the {\em two-norm} 
SVM formulations,
namely, Proximal SVM (PSVM)\cite{fung} and
Lagrangian SVM (LSVM)\cite{mangasarian}.
Cast as an optimization problem,
the objective functions associated with these particular SVM 
formulations 
are identical, while the constraints differ slightly:
\begin{equation}
  \min_{\hat{w}} \, \frac{1}{2}\|\hbd{w}\|_2^2 +
  \frac{\nu}{2}\|\bd{\xi}\|_2^2
  \spc \mbox{ s.t. } \spc
  \left\{ \begin{array}{lcll}
  \bd{Y} \left(\hbd{X}^T \hbd{w}\right) + \bd{\xi} & \geq & \ones{n} &
  \mbox{(LSVM)}\\
  \bd{Y} \left(\hbd{X}^T \hbd{w}\right) + \bd{\xi} & =    & \ones{n} &
  \mbox{(PSVM)}
  \end{array} \right.
  \label{eq:psvm_lsvm}
\end{equation}
where $\bd{\xi} \in \R^n$ is a vector of slack 
variables\cite{bertsekas}
and $\nu$ is a parameter which
controls the overall amount of misclassification 
violation\cite{cristianini}.
If the data points were linearly separable, 
the LSVM would require that all of the data points 
lie outside of the corridor defined by 
$g(\bd{w})=\hbd{x}^T \hbd{w}<1$.  Numerically, this 
is described by the set of constraints:
\begin{equation} 
  \bd{Y} \left(\hbd{X}^T \hbd{w}\right) \geq \ones{n}.
  \label{eq:sep_constraints}
\end{equation}
However, for data sets that are not linearly separable,
the constraints in Eq.(\ref{eq:sep_constraints}) 
are never simultaneously satisfied.  To allow for  
the likely possibility of data points lying on the wrong side of the 
separating hyperplane (or allowing for \emph{slackness} in the 
constraints), 
$\bd{\xi}$ and $\nu$ are introduced 
to allow for the violation of Eq.(\ref{eq:sep_constraints}).
In effect, one can view the LSVM relaxation of the inequality constraints
as a regularization mechanism  
which imposes a ``smoothness'' prior on the solution $\hbd{w}$.
The corresponding {\em dual} version of Eq.(\ref{eq:psvm_lsvm}),
obtained using the Karush-Kuhn-Tucker conditions of 
optimization theory\cite{bertsekas},
can then be expressed, after some algebraic manipulation, as the following
constrained LS problem,
\begin{equation}
  \mbox{solve} \spc (\bd{K} + \lambda^2 \bd{I}_n) \bdb = \bdy
  \spc \mbox{ s.t. } \spc
  \left\{ \begin{array}{rcll}
  \bd{Y} \bdb &\geq & \zeros{n} & \mbox{(LSVM)}\\
  \bdb & \in & \R^n             & \mbox{(PSVM)},
  \end{array} \right.
  \label{eq:kernel_psvm_lsvm}
\end{equation}
where $\lambda=1/\nu$.
For the PSVM, $\bdb$ is unconstrained in Eq.(\ref{eq:kernel_psvm_lsvm})
and hence is equivalent to KRR (this was first observed by
Agarwal\cite{agarwal}).
For the LSVM, the data points associated with the nonzero component values
of $\bdb$ are called the {\em support vectors}.  Geometrically, these data
points lie on or within the {\em canonical corridor} defined by 
$g(\bdx) = \hbd{x}^T \hbd{w} \leq 1$
and these points alone determine the orientation of the hyperplane 
since
$\hbd{w} = \hbd{X}\bdb = \sum_{\beta_i \neq 0} \beta_i \hbd{x}_i$.

\subsubsection{PLS}
\label{sec:pls_ff}
The Krylov subspace,
$\mathcal{K}_m(\bd{A},\bd{b}) =
  \mbox{span}\{\bd{b},\bd{A} \bd{b},\ldots,
               \bd{A}^{m-1} \bd{b} \}$,
is often associated with the conjugate gradient (CG)
algorithm applied to $\bd{Ax} = \bd{b}$ when $\bd{A}$ is
SPD\cite{trefethen}.
In solving the regression problem $\hbd{X}^T \hbd{w} = \bdy$,
Helland showed that the PLS-based weight vector $\hbd{w}$
lies in the Krylov subspace
$\mathcal{K}_m(\hbd{X}\hbd{X}^T,\hbd{X}\bd{y})$\cite{helland}.
If we express this Krylov subspace in terms of $\bd{K}$ and $\bdb$, 
we have
\begin{equation}
  \mathcal{K}_m(\hbd{X}\hbd{X}^T,\hbd{X}\bdy)
  = \mbox{span} \{ \hbd{X}\bdy,\hbd{X}\bd{K}\bdy,\ldots,\hbd{X}\bd{K}^{m-1}\bdy \}
  = \mbox{span} \{\hbd{X}\bdb\},
\end{equation}
where $\bdb \in \mathcal{K}_m(\bd{K},\bdy)$.  As a result, the modified
LS problem associated with PLS is
\[
  \bd{K} \bdb = \bdy, \,\, \bdb \in \mathcal{K}_m(\bd{K},\bdy).
\]
The filter factors can then
be expressed in terms of the Ritz polynomial, $p_m(\theta)$\cite{lingj},
\begin{equation}
    f_i \, = \, 1 - p_m(\sigma_i^2), \spc \spc
    p_m(\theta) = \prod_{i=1}^n
               \left(\frac{\theta_i(m)-\theta}
               {\theta_i(m)}\right),
    \label{eq:ff_pls}
\end{equation}
a consequence of the Lanczos tridiagonalization process inherent in 
the CG algorithm\cite{hansen}. At step $m$, the Lanczos 
tridiagonalization produces a tridiagonal matrix $\bd{T}_m$ by taking 
the QR factors of successive Krylov subspaces, $\bd{Q}_m \bd{R}_m := 
\mathcal{K}_m(\bd{K},\bdy)$, such that $\bd{T}_m = \bd{Q}_m^T \bd{K} 
\bd{Q}_m$\cite{trefethen}. The eigenvalues of $\bd{T}_m$, denoted by 
$\theta_i(m)$, are called the Ritz values. The filtering properties 
of $f_i$ are controlled by the convergence of $\theta_i^m$ to the 
nonzero eigenvalues of $\bd{K}$, \iie 
$\mbox{diag}(\Sigma^2)=\{\sigma_1^2,\ldots,\sigma_n^2\}$. This 
convergence, in turn, is controlled by the number of columns $m$ 
spanned by the Krylov subspace. If $\theta_i(m)$ has converged to 
$\sigma_i^2$, then $p_m(\theta_i(m))=0$ and $f_i=1$.  As $m$ 
approaches $n$ and if {\em all} of the Ritz values converge to their 
corresponding eigenvalues of $\bd{K}$, then all of the filter factors 
will approach one, and the PLS solution will coincide with the OLS 
solution. However, in practice, as $m$ increases, one often gets 
multiple copies for 
large values of $\theta_i(m)$ ({\em ghost values}) due to 
finite-precision arithmetic\cite{trefethen}. 
Hence, the Ritz values disproportionally converge 
to the largest eigenvalues of $\bd{K}$ as $m$ increases.  This 
results in a delayed or lack of convergence of $\theta_i(m)$
toward $\sigma_i^2$. 
Therefore, the PLS solution generally does not coincide in 
real computations with the OLS solution for 
$m=n$ since convergence to the smallest eigenvalues of $\bd{K}$ is 
rarely achieved.
As a result of the delayed convergence, an optimal or near optimal $m$ often satisfies 
$m \ll n$ for ill-conditioned coefficient matrices 
$\bd{K}$---see Fig.(\ref{fig:filfac}). 

\subsubsection{Nonstandard filter factor representations}
\label{sec:other_ff}
Filter factor representations can easily accommodate non-standard
filtering procedures.
Standard filtering procedures typically assume that, 
in the SVD expansion,
the terms corresponding to the  
largest singular values are the most important and should be kept,
while the terms corresponding to the smallest singular 
values should be
filtered out or damped.
This not need be the case with gene expression data.  
In Alter \etal\cite{alter}, 
the largest singular value and its corresponding right
and left singular vectors were removed in order to filter out the steady
state of yeast cell cycle expression. 
Bair \etal\cite{bair}
introduced a mixture model for outcome to accommodate other sources
of non-outcome variation in which the outcome class label
vector $\bdy$ was expressed as a linear combination of the first two
singular vectors:
$\bdy \approx \alpha \bd{u}_1 + (1-\alpha) \bd{u}_2 + 
\bd{\epsilon}_n$, 
$\alpha \in [0,1]$, where $\bd{\epsilon}_n$ is an 
$n \times 1$ error vector in which each component value 
is Gaussian 
with unit variance.
When $\alpha=1$ and $\alpha=0$, $\bdy$ is correlated with
$\bd{u}_1$ and $\bd{u}_2$,
respectively, since
$\bdb = \bd{UF\Sigma}^{-2}\bd{U}^T \bdy \approx
[f_1/\sigma_1^2] \alpha \bd{u}_1 + [f_2/\sigma_1^2] (1-\alpha) \bd{u}_2$.
Hence, for $\alpha = 1$, we have 
$f_1 \approx 1$ and $f_{i \neq 1} \approx 0$, while for
$\alpha = 1$, we have,
$f_2 \approx 1$ and $f_{i \neq 2} \approx 0$.

\subsection{Extent of regularization}
\begin{figure}[h]
\begin{center}
\caption{\fns Comparison of various filter factors
  representations for the training data in the  
  normalized \textbf{UNM Infant ALL/AML} data set 
  (see \S\ref{sec:data}).  The filter
  factors $f_i$ ($y$-axes) are plotted against the summation index $i$
  ($x$-axes).  The regularization parameters for TSVD, KRR and PLS are
  $k=\{5,36\}$ (the truncation parameters), $\lambda \in
  \{\sigma_{5},\sigma_{36}\}$ (the scalar multiples of the identity
  matrix, $\bd{I}_n$) and $m \in \{5,36\}$ (the number of columns
  spanning the Krylov subspace, $\mathcal{K}_m(\bd{K},\bdy)$).}
\includegraphics[width=13.0cm]{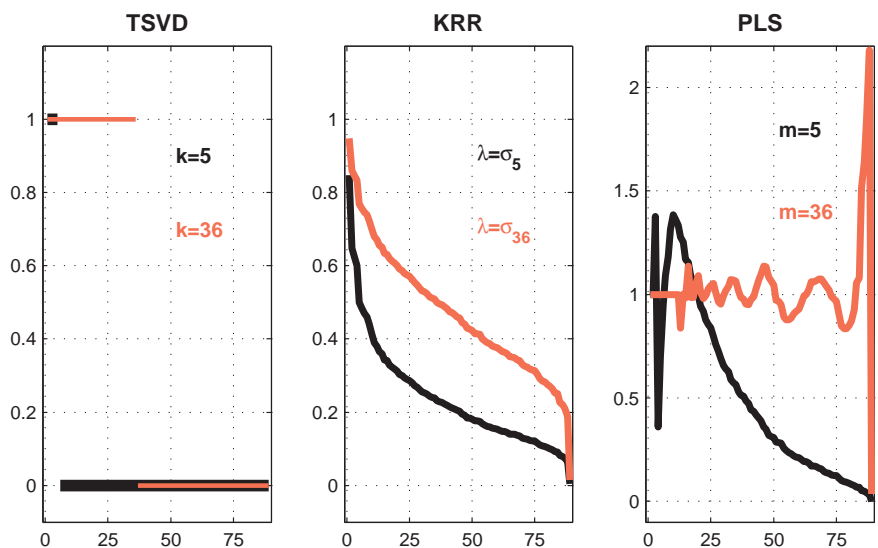}
    \label{fig:filfac}
\end{center}
\end{figure}

For TSVD, KRR and PLS, one must decide upon the value of 
$k$ (the number of singular values and vectors kept in the
SVD expansion of Eq.(\ref{eq:ols_soln})), $\lambda$ 
(the multiplicative factor of the identity matrix added
to $\bd{K}$ in Eq.(\ref{eq:krr})) and
$m$ (the number of columns spanned by the Krylov subspace
$\mathcal{K}_m$) using information obtained 
from the training data only.  
The choice of the regularization parameter dictates the 
values of the filter factors which, in turn, control the 
amount of smoothing the solution $\bdb$ will undergo. 
For TSVD and PLS, the regularization parameters 
$k$ and $m$ are the integer-valued entities 
$\{1,2\,\ldots,n\}$, and the smaller the regularization 
parameter, the greater the amount of smoothing. 
A regularization parameter 
of $n$ effectively amounts to an absence of regularization
since $f_i \approx O(1)$ for any $i \in \{1,2,\ldots,n\}$.
For KRR, however,
the regularization parameter $\lambda$ is nonnegative and
real-valued---the larger the value, 
the greater the amount of smoothing and vice-versa.
In particular, the effective range of regularization for
$\lambda$ lies in the interval $[0,\sigma_1]$ 
since $\lambda=0$ corresponds to no regularization 
($f_1=\cdots=f_n=1$) while $\lambda=\sigma_1$
roughly amounts  to keeping the first term 
of the SVD expansion 
($f_1=1/2 \approx O(1)$ and $f_i \approx O(\epsilon)$
for $i \geq 2$ and $0 < \epsilon \ll 1$).
Note that for TSVD and KRR, the filter factors lie within 
the interval $[0,1]$; see Fig.(\ref{fig:filfac}).  
From the 
point of view of regularization, this makes perfect sense since we
either want to keep an SVD expansion term 
($f_i \approx 1$) or suppress an SVD expansion term 
($f_i \approx 0$).  However, the filter factors for PLS
can lie outside of $[0,1]$ since   
$f_i = 1-p_m(\sigma_i^2)$ is a polynomial of degree $m$ and can 
assume large positive and negative values as $p_m(\theta)$ 
oscillates; see Fig.(\ref{fig:filfac}).  
Hence, the regularization effects
of PLS can be difficult to interpret.
Additional details of the regularizing effects of the CG 
algorithm in the context of PLS can found in the work of 
Lingj{\ae}rde and Christopherson\cite{lingj}. 

\subsection{Choice of filter factors and ill-posedness}
\label{sec:choice_of_filfac}
The choice of filter factors should be informed by the type of 
numerical ill-posedness exhibited by $\hbd{X}$. There are generally 
two types of numerical ill-posedness in the case of linear systems: 
\emph{rank-deficiency} and \emph{discrete ill-posedness}\cite{hansen}.  
Rank-deficiency is characterized by a large gap between $\sigma_k$ 
and $\sigma_{k+1}$ in the singular value spectrum of $\hbd{X}$ in 
which the last $n-k$ singular values are assumed to reflect spurious 
noise inherent in the data. In this case, $\txtsub{r}{eff} = k$ and 
the preferred treatment for handling such ill-posedness is the TSVD 
scheme: truncate the last $n-k$ terms in the SVD expansion.  On the 
other hand, discrete ill-posedness is characterized by a slow decay 
of the singular values in which there is no well-determined gap in 
the singular value spectrum.  For such problems, truncation of the 
SVD expansion may not lead to the best-regularized solution. If we 
truncate too early, then we may lose information, and if we include 
too many terms, then the solution can become unstable in the presence 
of noise. As a result, one may want to compromise by 
reweighting all of the terms such that terms with small singular 
values are damped to a greater degree than terms with larger singular 
values. In this case, the preferred remedy would be either KRR or 
PLS.  In this study, we are interested in whether the type of 
numerical ill-posedness exhibited by the cancer gene expression data 
should dictate the type of regularization scheme used and whether 
this leads to improvement in overall classification performance.
We now discuss our numerical results for several cancer gene expression 
data sets.

%
\section{Gene Expression Data Sets and Preprocessing}
\label{sec:data}
%

%
\subsection{Gene expression data sets}
Over the past several years, there have been 
a number of human gene expression
studies that have considered the problem of
 classification with respect to
tumor lineage and outcome.
This paper focuses on the following data sets:
\begin{itemize}
\item \textbf{UNM Infant ALL/AML}\cite{unm_infant}:
      In this study, 126 infant ($< 1$ year)
      samples were grouped according to their major precursor origins
      within acute leukemia:  acute lymphoblastic leukemia (ALL) and
      acute myeloid leukemia (AML).

\item \textbf{MIT-ALL/AML}\cite{golub}:
      In this study, 72 patient
      samples were grouped according to ALL or AML.

\item \textbf{UNM Pediatric ALL Outcome}\cite{unm_all}:
      254 pediatric patient samples ($< 18$ years of age)  
      with ALL were examined with respect to outcome.
      This data set includes distinct       
      tumor sublineages (34 T-cell ALL and 220 B-cell ALL patient samples) 
      and multiple molecular subtypes (the chromosomal aberrations that 
      give rise to certain leukemias).

\item \textbf{van't Veer Breast Cancer Outcome}\cite{veer}:  This data set
      concerns primary breast cancer outcome 
      (patients who remained metastasis-free for at least 
      five years versus those patients
      who developed distant metastases within five years)
      in a group of 95 patients selected for age ($< 55$ years) and a 
      clinical indication of favorable prognosis 
      (\iie lymph node negative status and a tumor diameter less than 5
      centimeters).
\end{itemize}
The class distribution of the patient samples across the cancer gene
expression data sets can be found in Table \ref{tab:classdist}.
The \textbf{UNM Infant ALL/AML} and the 
\textbf{UNM Pediatric ALL Outcome} data sets used the Affymetrix 
MAS 5.0 software\cite{affy} to generate expression levels for 12625
genes (actually cDNA probesets), while for the 
\textbf{MIT ALL/AML} data set, 
the Affymetrix MAS 4.0 software was used to generate expression 
values for 7129 genes.  For the 
\textbf{van't Veer Breast Cancer Outcome} data set, 
spotted gene arrays (Hu25K microarrays) containing 24,481 genes 
were used. 
\begin{table}[h]
\begin{center}
\caption{\fns Class distribution of patient samples across data sets.}
\begin{tabular}{|l|l|l|} \hline
{\fns Data Set} & {\fns Training Set} & {\fns Test Set}
\\ \hline \hline
{\fns \textbf{UNM Infant ALL/AML}}
& {\fns 54 ALL; 35 AML}
& {\fns 24 ALL; 13 AML} \\ \hline
{\fns \textbf{MIT ALL/AML}}
& {\fns 27 ALL; 11 AML}
& {\fns 20 ALL; 14 AML} \\ \hline
{\fns \textbf{UNM Pediatric ALL Outcome}}
& {\fns 73 REM; 94 FAIL}
& {\fns 39 REM; 48 FAIL} \\ \hline
{\fns \textbf{van't Veer Breast Cancer Outcome}}
& {\fns 44 REM; 34 FAIL}
& {\fns 7  REM; 12 FAIL} \\ \hline
\end{tabular}
\label{tab:classdist}
\end{center} \end{table}

These gene array data sets were 
chosen since we wanted to consider class 
prediction tasks based upon
tumor lineage and outcome.  In approximate terms, 
tumor lineage and outcome are
opposites 
with respect to the  
observed biological variability.  We expect 
tumor lineage to have a 
higher signal-to-noise ratio than outcome since tumor lineage 
dominates the fate of cancer cells. 
Conversely, outcome embodies the full biological, genomic and
environmental heterogeneity of an individual,
making it difficult to isolate a unique or universal favorable or 
unfavorable signature within the biological noise. 
Since hyperplane classifiers differ
in how they stabilize the solution in the presence of noise, certain
hyperplane classifiers may to be better-suited to handling class
prediction tasks that differ in signal-to-noise ratios.

\subsection{Gene selection}
In cancer gene expression studies, not all genes
change substantially in response to disease.  As a result,
a filtering procedure is often used to reduce the total of number
of genes, $d$, to a core subset in which the reduced number of genes,
denoted by $d'$, are differentially expressed with respect
to the class distinction of interest.
Here, a two-stage filtering procedure is
used.  First, genes are removed on the basis of
qualitative measures of gene detection.  Second,
highly discriminating genes are chosen using a
ranking procedure based upon 
weight-vector component magnitudes.

For the data sets using the Affymetrix platform,
all control genes/probesets were removed, \iie 
all probesets having the 
``AFFX" prefix in the probeset ID\cite{affy}.
In addition, any probeset that 
did not have at least one ``Present" \emph{call value}
(as determined by the 
Affymetrix 
MAS 5.0 statistical software) within the training set samples
was also removed.
A call value is an Affymetrix-specific qualitative measure of
detection:  it indicates whether a
gene was expressed (``present''), was not
expressed (``absent'') or was too close to call (``marginal'').
This typically reduces the initial number 
of probesets by a third or a quarter.
For the \textbf{van't Veer Breast Cancer Outcome} data set (the only
non-Affymetrix-based data set), two patients (samples 53 and 54)
had a significant number of missing values and these two samples 
were excluded from subsequent analysis (this was not done in
the study of van't Veer \etal\cite{veer}).    
A gene was then removed if there was at least one missing expression
value among all of the training set samples.
Note:  If one of the genes in the test set
had a missing value, 
then the missing value for that gene was replaced with the mean of the
normalized expression values for that gene in the training set;
see \S\ref{sec:normalization} for
data normalization details. 

For the second stage of the gene selection process, 
a modified version of the 
Recursive Feature Elimination (RFE) algorithm\cite{guyon} was used.
In the RFE
algorithm, an SVM (or any other hyperplane classifier for that matter) 
is trained using all of the genes from patient samples in the
training set.  The genes are then ranked according to their largest
weight-component magnitudes,
\[
  \{ |w_{i_1}|,\ldots,|w_{i_d}| \}, 
\]
where $\{i_1,\ldots,i_d\}$ are the sorting indices 
such that $|w_{i_1}|$ and $|w_{i_d}|$ are the largest and smallest
weight-component magnitudes, respectively.
The genes associated with the smallest
weight-component magnitudes are removed and the
SVM is then retrained using the remaining genes.
This process is repeated
iteratively until there are no more genes left.  
Here, we use KRR to generate the weight vector per iteration:
$\hbd{w}^{(k)} = \hbd{X} \bdb^{(k)}$ where
$\bdb^{(k)}$ is the solution to the linear system,
\[
  (\bd{K}^{(k)} + \lambda^{(k)} \bd{I}_n) \bdb^{(k)}= \bdy,
\]
$k$ is the RFE iteration and $\bd{K}^{(k)}$ is the
kernel matrix obtained using only the gene
subset obtained at iteration $k$.
In the previous work of 
Guyon \etal\cite{guyon}, the regularization 
parameter $\lambda$ was fixed across
all gene subsets.  This strategy tends to under-regularize 
(over-regularize) $\bd{K}^{(k)}$ when $k$ is small (large).
Here, we modify
the regularization parameter $\lambda^{(k)}$ to 
be the mean of the singular values of
$\bd{K}^{(k)}$ (or equivalently, the mean of the trace of $\bd{K}^{(k)}$).
This allows for {\em fast} computation (no
SVD of $\bd{K}^{(k)}$ is required) and {\em consistent}
filtering across RFE iterations.  This filtering also 
intentionally errs on the side of over-regularization 
since $\lambda^{(k)}$ is weighted toward the largest
singular values (otherwise, if $\lambda^{(k)}$ is too small, then
we will under-regularize $\bd{K}^{(k)}$ 
and overfit the training data).

In this study, the number of genes used in the 
hyperplane classifier is varied amongst the values in the 
set:
$d'=\{d\, \mbox{(all genes)},100,50,25,10\}$.
Our interest is not in finding the optimal subset of
discriminating genes, but in measuring qualitative changes of
class prediction performance as $d'$ is decreased.
Removing spurious ``gene noise'' is a common 
and valid reason given for using gene selection.
However, there are other more practical reasons for doing so.
Computationally, gene
selection greatly eases the numerical burden in, say,
computing the SVD of $\hbd{X}$.
Clinically, assays that require
relatively few gene expression levels to be measured
are more likely to
be adopted in a real-world setting.

\subsection{Data normalization}
\label{sec:normalization}
Expression values corresponding to different genes can differ
significantly in magnitude.
As a result, one often tries to transform and scale the data such
that the expression values across patients for a given
gene are of the same size.  First, the expression data was
$log$-transformed using the transformation,
$\bd{x}_{ij} \leftarrow sgn(x_{ij}) log_2(|x_{ij}|)$,
since the   
data tends to exhibit a log-normal distribution
for the {\em positive} values for a fixed gene
(note that for the \textbf{van't veer Breast
Cancer Outcome} data set, the publically-available data was
already log-transformed). 
Second, the expression values for a fixed gene across 
the training set samples
were normalized to mean zero and unit standard deviation (the 
expression values for a fixed gene within the test set were normalized 
using the mean and standard deviation from the training set).
Third, to avoid scaling problems when solving $\bd{K}\bdb = \bdy$,
both $\bd{K}$ and $\bd{y}$ were re-scaled so that their 
component elements were of commensurate size.  Following the
recommendation of Varah\cite{varah},
the left- and right-hand sides of $\bd{K}\bdb = \bdy$
were re-scaled such that
$\| \bd{K} \|_2 = \sigma_1^2 \approx O(1) \approx  \| \bdy \|_2^2$.
This was easily accomplished using the following re-scaling:
$\bd{\Sigma} \leftarrow \bd{\Sigma}/\sigma_1$
and $\bdy \leftarrow \bdy/\sqrt{n}$.

\subsection{Model Selection}
$N$-fold cross-validation was used to estimate the value of the
regularization parameter.  In this study, we used $N=10$.
For TSVD and PLS, the candidate values used for $k$ and
$m$, respectively, came from the set $\{1,\ldots,n\}$.
The candidate values for $\lambda$ came from the set of
singular values ($\lambda \in \{ \sigma_1,\ldots,\sigma_n \}$)
since, for Tikhonov and Franklin regularization, the 
effective range for $\lambda$ lies in the interval 
$[0,\sigma_1]$\cite{hansen}.
Due to the re-scaling of the data to obtain $\|\bd{K}\|_2^2 = 
\sigma_1^2=1$, the set of candidate values for $\lambda$
lie in the interval $[0,1]$.
Note that in this study, gene selection is performed
{\em separately on each cross-validation fold}.
Failure to do so will induce a gene selection bias that
yields overly optimistic cross-validation error
rates\cite{ambroise}.

%
\section{Results}
\label{sec:res}
%
%
\subsection{Software}
The software for the regularized least squares
hyperplane classifiers was written in MATLAB.  The MATLAB Regularization
Toolbox\cite{hansen_regtools} was used to calculate the CG-based filter
factors for the PLS algorithm.

\subsection{Classification performance}
\subsubsection{Performance averaged across many partitions} 
In gene expression studies, a single partition of the training and test set 
is often used for training and validation.  This specific partition is 
primarily chosen on the basis of case-control considerations
in order to mitigate any bias 
that may arise from the way the patient samples were collected.  However, 
drawing conclusions about classifier performance on the basis of a single 
training and test set partition can be misleading, especially when the 
sample size $n$ is small relative to the number of dimensions $d$. To avoid 
this dilemma, we measure classifier performance, on average, across 
many training and test set partitions. To maintain consistency across 
partitions, the number of positive and negative samples in the training and 
test set are kept the same. Overall classification performance was measured 
as an average across 1000 training and test set partitions.  The average 
number of misclassifications was computed as $\bar{M}=(\sum_{p=1}^{1000} 
M^p)/1000$ in which $M^p$ is computed as in 
Eq.(\ref{eq:misclass}) for the $p^{th}$
training or test set partition.  

\subsection{Trends in overall classification performance} 
\begin{figure}[h]
    \centering
    \caption{\fns For a given data set,
    the average number of test set misclassifications ($y$-axis) was
    plotted as a function of the number of genes 
    ($x$-axis).  
    Different colors correspond to
    different hyperplane classifiers:  \textbf{black} for
    $\mbox{SVM}_2$, the two-norm SVM
    of Eq.(\ref{eq:kernel_psvm_lsvm}), and 
    \textbf{red}, \textbf{blue} and \textbf{green} 
    for the SVD-based
    filter factor representations of TSVD, KRR and PLS, respectively.
    The squares correspond to the average
    number of misclassifications and the vertical bars 
    indicate the standard deviation.  The 
    numbers in parentheses next to the plot title 
    denote the number of misclassifcations
    that would be obtained by majority class prediction (MCP) and the total
    number of test samples used for that specific data set, respectively.}
    \includegraphics[width=13.0cm,height=9cm]
      {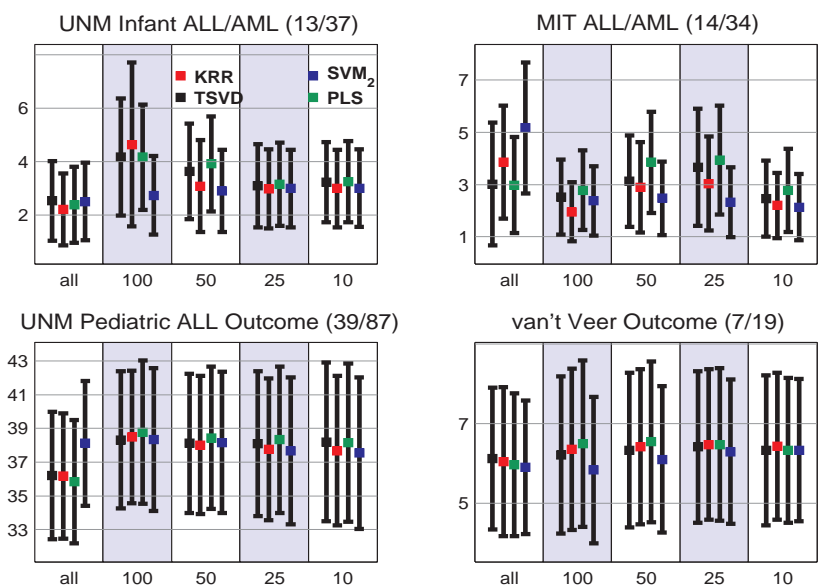} 
    \label{fig:err}
\end{figure}

For each data set, Fig.(\ref{fig:err}) displays the average number of 
misclassifications 
(and corresponding error bars given by the standard deviation)
for four different hyperplane classifiers and five distinct feature-set 
choices (number of genes).
Three out of the four classifiers correspond to the 
standard least squares techniques
(TSVD, KRR and PLS) and the fourth classifer corresponds to the
two-norm SVM formulation of 
Eq.(\ref{eq:kernel_psvm_lsvm}), hereafter denoted as $\mbox{SVM}_2$. 
The 
average number of misclassifications for each data set is compared to 
the number of misclassifications that would be obtained by majority class 
prediction (MCP), where a prediction for a new sample is 
determined solely by siding 
with the class in the test set that has the most members (this is not known 
\emph{a priori}).

In Fig.(\ref{fig:err}), the subplots in 
rows one and two correspond to the data sets associated with 
tumor lineage and outcome, respectively.
As expected, class prediction tasks related to tumor lineage are
`easy' 
(as evidenced by the low number of misclassification and small 
standard deviations)
since tumor lineage dominates the expression 
behavior. For outcome, however,
the results were only slightly
better than what would be obtained by MCP.
On average, the choice of regularization scheme did not impact class
prediction performance.  The simplest of the 
filter factor strategies,
the TSVD scheme, sufficed in most instances.

Somewhat surprisingly, gene selection 
did not improve performance.  In fact, gene selection
slightly degraded performance for outcome prediction as the number of
genes was decreased.  For tumor lineage, only KRR and $\mbox{SVM}_2$ 
slightly benefited from gene selection for $d'<100$.  This is not too 
surprising since KRR and  $\mbox{SVM}_2$ share the same regularization 
mechanism, \iie the Franklin regularization scheme.
Particularly striking was  
the poor performance of the least squares
classifiers relative to $\mbox{SVM}_2$ for $d'=100$ in the 
\textbf{UNM Infant ALL/AML}
data set.  In this instance, the average number of support 
vectors was 83 patient samples out of a total of 89 training set samples.
Since every patient expression vector is a support vector for
the regularized least squares classifiers,
``outlier'' patient expressions vectors (perhaps the 
6 non-support vectors not used by $\mbox{SVM}_2$)
can possibly 
tilt the weight vector $\hbd{w}$ in such a way that 
does not generalize well to new data points. 
Aside from this one instance,
the regularized least squares classifiers, 
compared against $\mbox{SVM}_2$, 
were competitive in terms of prediction accuracy, considerably
simpler to implement, and faster in terms of execution time. 
Since TSVD sufficed in most instances,
we will now examine its behavior in detail. 

%
\subsection{Spectral coefficients}
%
The comparable classification performance of TSVD 
relative to its other, more complicated, filter factor rivals 
suggests that the type of numerical ill-posedness exhibited 
by the LS problems derived from the cancer expression data 
sets in this study tends towards rank-deficiency as opposed 
to discrete ill-posedness. However, classification performance 
alone is not the only benchmark measure that can be used to
assess whether LS problems, derived from cancer expression data,
tend toward rank-deficiency in general.  
The \emph{spectral coefficients}
in Eq.(\ref{eq:ff_soln}) can also be used to indicate the type of 
numerical ill-posedness.  The spectral coefficients consist of 
the singular values ($\{\sigma_1,\ldots,\sigma_n \}$) and the 
Fourier coefficients 
($\{ |\bd{u}_i^T \bd{y}|,\ldots,|\bd{u}^T \bd{y}| \}$).  
Both of these coefficients decrease in value as $i$ increases and, 
by examining their rates of decay, they shed insight 
into the type of numerical ill-posedness of the 
LS problem.  
In addition, one can use the decay behavior of these coefficients 
in estimating the truncation parameter for TSVD.
The utility of these coefficients in determining 
the type of numerical ill-posedness and the TSVD truncation
parameter for hyperplane classification of cancer gene 
expression data is discussed next.

\subsubsection{Gaps in the singular values}
\begin{figure}[ht]
    \centering
    \caption{\fns For a given data set, the $y$-axes denote
    averaged singular values (first row),
    averaged Fourier coefficients (second row) and
    averaged number of misclassifications for the TSVD
    method (third row). The $x$-axes denote the
    summation index $i$ or the truncation parameter
    $k$ in the case of the averaged number of misclassifications.
    Different colors correspond to the different numbers of
    genes used in the analysis (see legend).}
    \includegraphics[width=13.0cm,height=9cm]
      {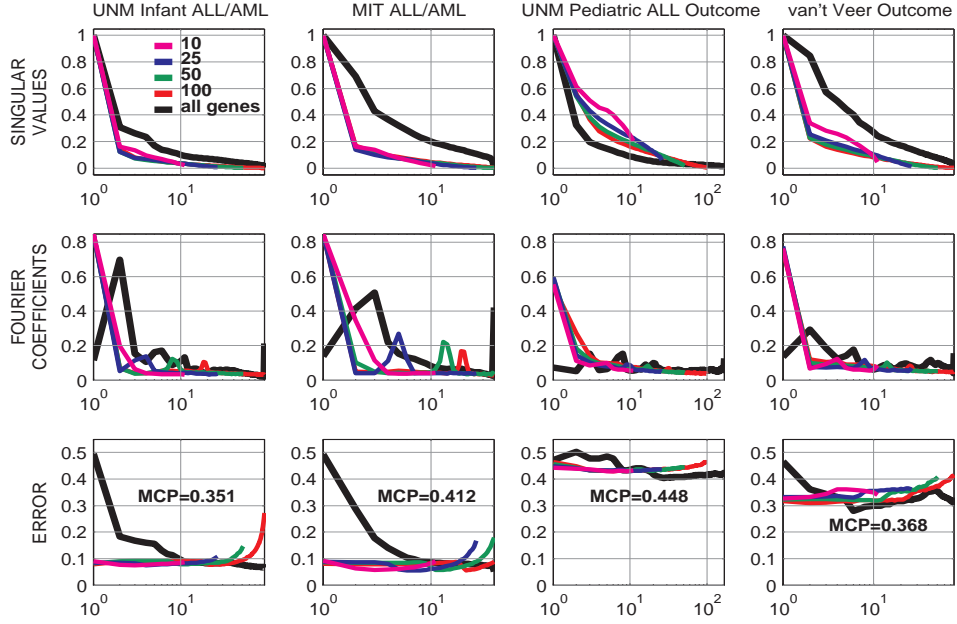}
    \label{fig:spec}
\end{figure}

As mentioned previously
in Section \ref{sec:choice_of_filfac},
rank-deficiency is characterized by a large gap between
singular values $\sigma_k$ and $\sigma_{k+1}$ such that
$f_i=1$ for $i \in \{1,\ldots,k\}$ and $f_i=0$ otherwise.
In such a case, a general rule of thumb would be that the index 
of where the gap occurs (\iie $k$) corresponds to the  
truncation parameter for TSVD. 
We now want to see if there is a large gap 
in the averaged singular value spectrum across all training set
partitions such that
$\bar{\sigma}_i = (\sum_{p=1}^{1000} (\sigma_i^p)^2)/1000$, 
$i=1,\ldots,n$. 
In Fig.(\ref{fig:spec}) (top row), 
we have plotted the singular values 
for varying numbers of genes in each of the 
four datasets.
The averaged rate 
of decay when using all genes (the solid black line), with the 
exception of the \textbf{UNM Pediatric ALL Outcome} data set, 
is faster than when using fewer genes.
This is to be expected since gene selection should, in 
principle, remove gene noise by keeping only the most {\em class-specific} 
genes.  For the 
\textbf{UNM Pediatric ALL Outcome} data set
the opposite is true: the rate of decay 
decreases as $d'$ decreases.  What could cause this rank inflation as $d'$ 
decreases in this instance? One possible answer to this question lies in the 
heterogeneity of this data set. The 
\textbf{UNM Pediatric ALL Outcome} 
data set can be grouped 
according to the two major lymphocyte subtypes: 
B-cells (34 samples) and T-cells (220 samples).
They may also be characterized by molecular subtype or chromosomal 
abnormality (30 t(12;21), 30 t(1;19), 14 t(9;22), 20 t(4;11) and 29 
hyperdiploid samples, with 131 `other' molecular subtype samples). Since tumor 
lineage and chromosomal abnormalities determine the fate of cancer cells, 
these class distinctions dominate the observed expression behavior
in terms of signal recovery.  
Classifying expression on the basis of these 
class distinctions is expected be relatively easy since the gene expression 
profiles corresponding to different cell or chromosomal subtypes have 
relatively large signal-to-noise ratios. 
Indeed, this is borne out in tests\cite{kang,willman}.
Hence, if one is interested in 
outcome signal recovery, then one has to contend with 
the relative weakness of the outcome signal compared to the 
dominant biological signals of tumor sublineages 
and molecular subtypes. Furthermore, the gene selection process for outcome 
may be confounded since the signal-to-noise ratio is low: in essence,
one is 
attempting to find outcome-specific genes by training on noise. 
Unfortunately, we believe that this hypothesis cannot be adequately addressed using real 
expression data.  Thus, in order to assess whether gene selection 
techniques 
actually find 
the class-specific genes of interest when there are many 
competing sources of 
biological variability, a simulation approach will most likely be 
necessary. 

%
\subsubsection{Fourier coefficients}
For the four gene expression data sets examined, 
the decay of the Fourier coefficients proved
more illuminating than the singular value decay
in terms of determining the type of
ill-conditioning.  In the presence of noise, 
the Fourier coefficients $|\bd{u}_i^T \bdy|$
will level off at a base-line
level of background noise determined by the errors in the coefficient
matrix $\bd{K}$ for $1 \leq i_0 < i  \leq n$\cite{hansen}. 
As a result, the SVD expansion terms for $i>i_0$ 
should, in principle, be filtered out since these terms 
are contaminated by
noise and will dominate the OLS solution.  
Under this criterion, $i_0$ can be used as a proxy
for the TSVD truncation parameter, \iie we set $k \approx i_0$.

The second row of Fig.(\ref{fig:spec}) displays the averaged Fourier
coefficients, 
$\bar{c}_i = (\sum_{p=1}^{1000} |(\bd{v}_i^p)^T y^p|)/1000$,
$i=1,\ldots,n$.
Note that as the number of genes decreases from $d'=d$ 
to $d'=\{100,50,25,10\}$ across all four data sets, there is a qualitative
change in the decay of $\bar{c}_i$. When $d'=d$, the
decay of $\bar{c}_i$ (given by the solid black line) slowly oscillates
toward zero and $i_0 \approx r=\mbox{min}\{n-1,d'\}$ ($r=n-1$ or
nearly full rank when $n<d$ and $r=d$ when $n>d$).  This indicates
that all terms
in the SVD expansion should be kept and the OLS solution would suffice in
effectively classifying the expression data.  This was experimentally borne
out by examining the average number of misclassifications as a function of
the truncation parameter, 
denoted by $\bar{M}_k=(\sum_{p=1}^{1000} M_k^p)/1000$, 
where 
\begin{equation}
     M_k^p = \sum_{j=1}^N I [t_j^p ( (\, \hbd{z}_j^p)^T \hbd{w}^{(k,p)}\, )]
     \spc 
     \left(
     \hbd{w}^{(k,p)}=\sum_{i=1}^k \frac{(\bd{v}_i^p)^T \bdy^p}
                                       {(\sigma_i^p)^2} 
                                       f_i^p \bd{u}_i^p
     \right) 
\end{equation}
is the number of misclassification per truncation parameter $k$ and 
partition $p$. 
In the third row of Fig.(\ref{fig:spec}), $\bar{M}^k$ is 
plotted against $k$.
When $k \approx r$ and $d'=d$,
the minimum value or a near minimum value for
$\bar{M}^k$ was obtained for tumor lineage and outcome, respectively.

When using fewer genes ($d'<d$), the decay of
$\bar{c}_i$ (the colored lines)
descends quickly to 0 such that $i_0 \approx 1$
(the exception was the \textbf{UNM Pediatric ALL Outcome} 
data set).
In this case, $i_0 \approx 1$ implies that a low-rank
approximation for $\bd{K}$ 
suffices for effective classification.
Again, this was experimentally borne out since
$k \approx 1$ was the truncation parameter 
that minimized
$\bar{M}_k$ (in good agreement with the 
effective resolution limit of
$i_0 \approx 1$).
Notice that for the \textbf{UNM Pediatric ALL Outcome} data set,
the Fourier coefficient decay was not as rapid as with
the other data sets (\iie $i_0 \gg 1$).
Again, there was good experimental agreement between the value of 
$i_0$ and the truncation parameter
($k \gg 1$) that minimized $\bar{M}_k$ in the third row of 
Fig.(\ref{fig:spec}).
While $i_0$ does appear to be a reasonable proxy for a near-optimal
TSVD truncation parameter, caution is still advised in the case
of outcome prediction since the best misclassification results were, on
average, barely below MCP.

%
\section{Conclusion and Future Work}
\label{sec:concl}
%

In this study, we have shown that many
popular least squares techniques used for linear discrimination
can easily be united under the framework of an SVD-based
filter factor representation.
Using the philosophy of Occam's razor as a guide, and based upon the 
four data sets examined,
the TSVD regularization
scheme emerged as the preferred hyperplane regularization strategy 
since its
performance was comparable, on average, to $\mbox{SVM}_2$ 
and to the
other more complicated filter factor strategies.
The classification performance of TSVD, coupled with the 
decay behavior of the Fourier coefficients in 
Fig.(\ref{fig:spec}),
lends credence to the observation that
linear inverse problems associated with the hyperplane classification
of gene expression data tend toward rank-deficiency as opposed to
discrete ill-posedness, especially when using fewer genes.
The spectral coefficients, in particular the Fourier coefficients,
act as useful numerical diagnostics, indicating when
signal recovery for outcome or tumor subclass  is possible.

Gene selection did not necessarily lower the number of 
misclassifications. Indeed, for outcome, the best performance was 
often achieved using all $d$ genes. 
One possible explanation is that the genes responsible 
for distinguishing differences in outcome are dominated by a significant 
number of differentially expressed genes responsible for the major sources 
of \emph{upstream} biological variation, \ieg tumor subclass distinctions or 
molecular subtype in the case of leukemia.  
If this is the case, signal 
recovery for outcome is likely to be 
severely hampered because the major sources of 
variation (biological and experimental) can inflate the singular 
value spectrum to such an extent that the only signals that remain above the 
noise background are those which dominate the expression behavior.
The testing of this hypothesis 
will require extensive gene array simulations and the modeling of cancer gene 
expression as a superposition of tumor subclass, outcome and experimental 
set effect signals, with class-specific sets of genes responsible for 
distinguishing class subtypes within each signal. This is a subject for 
future study.  Alternatively, 
one can restrict the patient sample size to create data subsets that are less 
heterogeneous in terms of dominant sources of variation, \iie  gene 
selection
within samples restricted to B-cell or T-cell acute lymphoblastic leukemias only.
However, one then pays the price of working with data sets that are 
of extremely small size.
In general, gene selection, 
by preserving the most class-specific genes,
allows for very aggressive regularization  
by making the kernel matrix more 
rank-deficient---as indicated by the 
rapid decay of the Fourier coefficients 
from $d'=\{\mbox{all genes}\}$ to 
$d' \leq 100$ genes.  This results in the ability to construct 
low-rank approximations of the kernel matrix $\bd{K}$.

The main reason for the success of SVD as an analysis tool is 
that it provides a new coordinate system in which the coefficient
matrix $\bd{K}$ becomes 
diagonal 
(\iie $\bd{\Sigma}=\mbox{diag}(\sigma_1,\ldots,\sigma_n)$). 
As a consequence, SVD is {\em rank-revealing}: the largest singular
values and vectors capture the relevant solution information by eliminating or
downweighting noise terms (using filter factors)
in the SVD expansion associated with the smallest 
singular values.
However, there are alternatives to the SVD, \iie there exist 
other rank-revealing matrix factorization
techniques such as rank-revealing QR or UTV 
factorizations\cite{hansen} 
that can also reliably solve  
rank-deficient LS problems.  Since TSVD works well with
regularized LS techniques 
in classifying gene expression data, it is likely that 
these alternative matrix factorizations will work just as 
well.
Moreover, and unlike SVD, these 
alternative matrix factorizations permit 
for the efficient 
updating of rows and columns when they are appended or 
deleted from $\bd{K}$\cite{hansen}, a common scenario in gene 
expression studies since patients and genes are 
routinely added or deleted from analysis as new clinical 
and experimental 
information becomes available.
Investigations of such alternative factorization approaches
in hyperplane classifers is currently in progress.

  \section*{Acknowledgements}
This work was supported in part by grants from the D.H.H.S. 
National Institute of Health NCI CA88361 and NCI CA32102, 
the W.M. Keck Foundation, and funds from the 
Dedicated Health Research Fund of the State of New Mexico.
This work was also supported in part by NSF Grant DMS-0306285.  
Any conclusions or recommendations expressed in this paper are 
those of the author and do not necessarily reflect the views of the NSF.
Erik Andries would like to acknowledge the UNM 
Center for High Performance Computing (CHPC)
for graduate fellowship support.  The authors also thank the CHPC
for providing the computational resources used in this work.

\end{document}